\documentstyle[12pt]{article}  
\begin{document}  
\centerline{\Large \bf On the supersymmetric partition function}  
\centerline{\Large \bf  in QCD-inspired random matrix models}  
\vskip 0.5cm  
\centerline{ \large \bf Yan V Fyodorov$^{1,2}$ and Gernot Akemann$^3$} 
\vskip 0.3cm \centerline{$^1$Department of Mathematical Sciences, 
Brunel University}   
\centerline{Uxbridge, UB8 3PH, United Kingdom} 
\centerline{$^2$Petersburg Nuclear Physics Institute, Gatchina
188350, Russia}   
\centerline{$^3$ Service de Physique Th\'eorique, CEA/DSM/SPhT}  
\centerline{Unit\'e de recherche associ\'ee au CNRS}  
\centerline{F-91191 Gif-sur-Yvette Cedex, France}  
\vskip 0.3cm   

\begin{abstract} We show that the expression for the supersymmetric
partition function of the chiral Unitary (Laguerre) Ensemble
conjectured recently by Splittorff and Verbaarschot \cite{1} follows
from the general expression derived recently by Fyodorov and Strahov
\cite{FS2}.  
\end{abstract}   

A class of random matrices that has attracted a considerable
attention recently
\cite{chirQCD,chirdis,chirmat,JSV1,GW,JNZ,BH,Ak,DV,my} is the
so-called {\it chiral} (Gaussian) Unitary Ensemble (chGUE), also
known as the Laguerre ensemble. The
corresponding matrices are of the form
$\hat{D}=\left(\begin{array}{cc}{\bf 0}&\hat{W}\\ \hat{W}^{\dagger}&
{\bf 0} \end{array}\right)$, where $\hat{W}$ stands for a complex
matrix, with $\hat{W}^{\dagger}$ being its Hermitian conjugate.  The
off-diagonal block structure is characteristic for systems with
chiral symmetry. The chiral ensemble was introduced to provide a
background for calculating the universal part of the microscopic
level density for the Euclidian QCD Dirac operator, see \cite{Ver}
and references therein. Independently and simultaneously it was
realised that the same chiral ensemble is describing a new group
structure associated with scattering in disordered mesoscopic wires
\cite{chirdis}.  One of the main objects of interest in QCD is the
so-called Euclidean partition function used to describe a system of
quarks characterized by $n_f$ flavors and quark masses $m_f$
interacting with the Yang-Mills gauge fields. At the level of Random
Matrix Theory the true partition function is replaced by the matrix
integral: 
\begin{equation}\label{partf} Z_{n_f}(\hat{M}_f)=\int {\cal
D}\hat{W}\prod_{k=1}^{n_f}\det\{i\hat{D}+m^{(k)}_f{\bf 1}_{2N}\}
e^{-N \mbox{\small Tr}V\left(\hat{W}^{\dagger}\hat{W}\right)}
\end{equation} 
where $\hat{M}_f=\mbox{diag}\left(m^{(1)}_{f},...,m^{(n_f)}_{f}\right)$ and
$V(z)$ is a suitable potential. Here the integration over
complex $\hat{W}$ replaces the functional integral over gauge field
configurations \cite{Ver}. Then the calculation of the partition
function amounts to performing the ensemble average of the product of
characteristic polynomials of $i\hat{D}$ over the probability density
$P(W)\propto e^{-N \mbox{\small
Tr}V\left(\hat{W}^{\dagger}\hat{W}\right)}$. In the general case of
non-zero topological charge $\nu>0$ the matrices $\hat{W}$ have to be
chosen rectangular of size $N\times (N+\nu)$ \cite{Ver}.  For
simplicity one may choose the probability distribution to be Gaussian
as defined by the formula: $d{\cal P}(W)\propto
d\hat{W}d\hat{W}^{\dagger}\exp{-\left[N\mbox{Tr}
\hat{W}^{\dagger}\hat{W}\right]}$.  

The characteristic feature of the
chiral ensemble is the presence of a particular point $\lambda=0$ in
the spectrum, also called the "hard edge" \cite{chirmat}. The
eigenvalues of chiral matrices appear in pairs $\pm\lambda
_k\,\,,\,\, k=1,...,N$. Far from the hard edge the statistics of
eigenvalues is practically the same as for usual GUE matrices without
chiral structure, but in the vicinity of the edge eigenvalues behave
very differently.  

Let ${\cal Z}_N[m]$ be the following spectral
determinant (characteristic polynomial of $i\hat{D}$): 
\begin{equation} 
{\cal Z}_N[m]=\det{\left(m^2{\bf
1}_N+\hat{W}^{\dagger}\hat{W}\right)}\ . 
\end{equation}  
and let us consider a more general (supersymmetric) partition function for
the chGUE defined as 
\begin{equation}
\label{partf1} {\cal K}(\hat{M_f},\hat{M_b})=
\left\langle \frac{
\prod\limits_{j=1}^{L}{\mathcal{Z}}_N\left[m^{(j)}_f\right]}
{\prod\limits_{j=1}^{M}{\mathcal{Z}}_N\left[m^{(j)}_b\right]}
\right\rangle_{W}\end{equation} where
$\hat{M}_f=\mbox{diag}\left(m^{(1)}_f,...,m^{(L)}_f\right)\,\,,\,\,
\hat{M}_b=\mbox{diag}\left(m^{(1)}_b,...,m^{(M)}_b\right)$.  This
correlation function contains much more information on spectra of
chiral matrices than the partition function (\ref{partf}) since it
involves both product and ratios of the characteristic polynomials.
Many efforts were spent on developing methods allowing one to
calculate particular cases of such a general supersymmetric partition
(or correlation) function \cite{DV,my}. In particular, the case $\nu=0$
 was completely solved
in \cite{FS1} by a variant of the supermatrix method \cite{Efetov,my}
augmented with a generalization of an Itzykson-Zuber type integrals
\cite{JSV1,GW} to integration over non-compact group manifolds. In
the recent paper \cite{1} Splittorff and Verbaarschot conjectured the
result for arbitrary integer $\nu>0$ in the microscopic (sometimes
also called "chiral") large-$N$ limit: $N\to\infty$ such that
$\hat{X}_{b,f}=2N\hat{M}_{b,f}$ is finite. The authors of \cite{1} used the
advanced version of the replica method suggested recently by
Kanzieper \cite{kanzieper}.  The final result is given in terms of a
determinant containing modified Bessel functions $I_l(z)$ ("compact
integrals") and their noncompact partners - Macdonald functions
$K_l(z)$.  The goal of the present Letter is to show that the case
$\nu\ne 0$ considered in \cite{1} in fact follows from a very general
expression derived in the recent paper \cite{FS2}. The demonstration
of this fact also provides a natural explanation why both compact and
non-compact integrals must appear on equal basis.  

The eigenvalues $x_1,....,x_N$ of the $N\times N$ positive definite matrix
$H=\hat{W}^{\dagger}\hat{W}$ are known to be distributed according to
the Laguerre ensemble density function 
\begin{equation}\label{Lag}
{\cal P}(x_1,...,x_N)\propto \Delta^2(\hat{X}) \prod_{i=1}^N
w_{\nu}(x_i) 
\end{equation} 
where $w_{\nu}(x)=x^{\nu}e^{-Nx}$ and
$\Delta(\hat{X})=\prod_{i>j}(x_i-x_j)$.   Note that the spectral
determinant ${\cal Z}_N(m)=(-1)^N\prod_{i=1}^N
\left[(-m^2)-x_i\right]$ is just the characteristic polynomial of
matrices $H$ from the Laguerre ensemble taken at negative real values
of the spectral parameter.  As is proved in the paper \cite{FS2} one
can express the general correlation function of the characteristic
polynomials for an arbitrary unitary invariant ensemble of $\beta=2$
symmetry class in terms of a $(M+L)$-sized determinant.  The main
building blocks of that determinant are (monic) orthogonal
polynomials $\pi_n(x)=x^n+...$ satisfying 
\begin{eqnarray}\label{1}
\int_{D}dx\, w(x) \pi_n(x)\pi_m(x)=\delta_{nm}c_n^2 
\end{eqnarray}
where $w(x)$ is a general weight function, $c_n$ are normalization
constants and $D$ is the corresponding interval of orthogonality.  A
novel feature revealed in \cite{FS2} is that for $M>0$ such a
determinant structure contains the {\it Cauchy transforms} of the
orthogonal polynomials 
\begin{eqnarray}\label{2}
h_n(\epsilon)=\frac{1}{2\pi i}\int_{D} dx
\frac{w(x)}{x-\epsilon}\pi_n(x) 
\end{eqnarray} 
alongside with the orthogonal polynomials themselves.  
For them to be well defined we need to have $Im(\epsilon)\neq 0$.  

Actually, the partition function
Eq.(\ref{partf1}) is given by \cite{FS2}:
\begin{eqnarray}\label{GeneralFormula} &&{\cal K}(\hat{M_f},
\hat{M_b})\propto \frac{1}{\triangle(\hat{M}_b^2)
\triangle(\hat{M}_f^2)}\\ \nonumber &&\times \mbox{det}\left|
\begin{array}{cccc} h_{N-M}\left(-[m^{(1)}_b]^2\right) &
h_{N-M+1}\left(-[m^{(1)}_b]^2\right) & \ldots &
h_{N+L-1}\left(-[m^{(1)}_b]^2\right) \\ \vdots &\vdots & &\vdots \\
h_{N-M}\left(-[m^{(M)}_b]^2\right) & h_{N-M+1}\left(-[m^{(M)}_b]^2\right)
& \ldots & h_{N+L-1}\left(-[m^{(M)}_b]^2\right) \\
\pi_{N-M}\left(-[m^{(1)}_f]^2\right) &
\pi_{N-M+1}\left(-[m^{(1)}_f]^2\right) & \ldots &
\pi_{N+L-1}\left(-[m^{(1)}_f]^2\right) \\ \vdots &\vdots & &\vdots \\
\pi_{N-M}\left(-[m^{(L)}_f]^2\right) &
\pi_{N-M+1}\left(-[m^{(L)}_f]^2\right) & \ldots &
\pi_{N+L-1}\left(-[m^{(L)}_f]^2\right) \end{array}\right|
.\end{eqnarray} 
For the Laguerre ensemble of matrices $H$ with
positive eigenvalues $\hat{X}=\mbox{diag}(x_1,...,x_N)$, the weight
function is just $w_{\nu}(x)=x^{\nu}e^{-Nx}$ , the domain is
$D=\left[0\le x<\infty\right]$ and the monic polynomials are
$\pi_{n}(x)=\frac{(-1)^n}{N^n}n!\,L_n^{\nu}(xN)$, with
$L_n^{\nu}(xN)$ being the standard Laguerre polynomials. Here $\nu$
can be taken real valued with $\nu>-1$.  To calculate the Cauchy
transform Eq.(\ref{2}) we exploit a well-known integral
representation for the Laguerre polynomials containing the Bessel
function $J_{\nu}\left(x\right)$: 
\begin{eqnarray}\label{3}
\pi_n(x)=\frac{(-1)^n}{N^{n+\nu/2}}\,e^{Nx}x^{-\nu/2}
\int_0^{\infty}dt\, e^{-t}t^{n+\nu/2}J_{\nu}\left(2\sqrt{Nxt}\right)
\, . 
\end{eqnarray} 
Let us consider, for definiteness,
$Im(\epsilon)>0$ and further employ the integral representation
\begin{eqnarray}\label{4}
\frac{1}{x-\epsilon}=i\,\int_0^{\infty}d\tau e^{-i\tau(x-\epsilon)}\,
\end{eqnarray} 
Then replacing $\pi_n(x)$ in (\ref{2}) by (\ref{3})
and $1/(x-\epsilon)$ by (\ref{4}) we easily perform the integration
over $x$ first, then integrate over $\tau$ and arrive at the
following representation (cf.Eq.(\ref{3})): 
\begin{eqnarray}\label{5}
h_n(\epsilon)=\frac{(-1)^n}{2N^{n+\nu/2}}\epsilon^{\nu/2}
\int_0^{\infty}dt\,
e^{-t}t^{n+\nu/2}H^{(1)}_{\nu}\left(2\sqrt{N\epsilon t}\right)\ .
\end{eqnarray} 
Here $H^{(1)}_{\nu}\left(z\right)$ is the Hankel
function of the first order.  Being actually interested in
analytically continued values of $\pi_n(x),h_n(x)$ for the region
$x=-m^2<0$ we introduce the modified Bessel and Macdonald functions
according to $I_\nu(z)=e^{-i\pi\nu/2} J_{\nu}\left(iz\right)$ and
$K_\nu(z)=\frac{i\pi}{2}e^{i\pi\nu/2} H^{(1)}_{\nu}\left(iz\right)$.
We then have 
\begin{eqnarray} \label{intrep}
&&\pi_n(-m^2)=\frac{(-1)^n}{N^{n+\nu/2}}\,e^{-N\,m^2}m^{-\nu}
\int_0^{\infty}dt\, e^{-t}t^{n+\nu/2}I_{\nu}\left(2m\sqrt{Nt}\right)
\\ && h_n(-m^2)=\frac{(-1)^n}{N^{n+\nu/2}}\frac{m^{\nu}}{i\pi}
\int_0^{\infty}dt\, e^{-t}t^{n+\nu/2}K_{\nu}\left(2m\sqrt{Nt}\right)
\end{eqnarray} 
Substituting such representations into the expression
Eq.(\ref{GeneralFormula}) it is easy to satisfy oneself that the
right-hand side can be rewritten as $(M+L)$-fold integral
\begin{eqnarray} &&{\cal K}(\hat{M}_f,\hat{M}_b)\propto
\frac{1}{\triangle(\hat{M}_b^2)\triangle(\hat{M}_f^2)}
e^{-N\sum_{j=1}^L [m^{(j)}_f]^2}
\left[\frac{\det{M_b}}{\det{M_f}}\right]^{\nu} \\ \nonumber && \times
\int_{t_i>0} d\hat{t}\det(\hat{t})^{N-M}
\Delta(\hat{t})e^{-N\mbox{\tiny Tr}\hat{t}} \prod_{l=1}^{L}\left[
t_l^{\nu/2}I_{\nu}(2m_f^{(l)}\,N\sqrt{t_l})\right]
\prod_{l=L+1}^{L+M}\left[t_l^{\nu/2}
K_{\nu}(2m_{b}^{(l-L)}\,N\sqrt{t_l})\right].  
\end{eqnarray} 
Here $\hat{t}>0$ is a diagonal matrix of the size $M+L$ with entries
$t_1,...,t_{M+L}$, and we rescaled $t\to Nt$.  Such an equation
generalizes the integral representation Eqs.(28)-(29) from \cite{FS1}
to nonzero values of $\nu$.  It is valid for any integer $N,L,M$.
The chiral limit $N\to\infty$ can be performed exactly along the same
lines as in \cite{FS1} and the result emerging is the one conjectured
by Splittorff and Verbaarschot \cite{1}, \cite{note}:
\begin{equation} {\cal K}(X_f,X_b)\propto
\left[\frac{\det{X_b}}{\det{X_f}}\right]^{\nu}\frac{1}{\triangle(X_b^2)
\triangle(X_f^2)} \det\left[X_i^{j-1}{\cal
J}_{\nu+j-1}(X_i)\right]_{i,j=1,\ldots,M+L}. 
\end{equation} 
Here $X_f=X_{\{i=1,\ldots,L\}}$ and $X_b=X_{\{i=L+1,\ldots,L+M\}}$ denote
the rescaled fermionic and bosonic masses respectively as well as
${\cal J}_i=I_i$ for $i=1,\ldots,L$ and ${\cal J}_i=K_i$ for
$i=L+1,\ldots,L+M$.  Note that the presence of "compact" (modified
Bessel) and "non-compact" (Macdonald) functions in the final
expressions is a direct consequence of the presence of both
orthogonal polynomials and their non-polynomial partners (Cauchy
transforms) in the determinantal representation.  One may also wish
to consider a more general type of potentials $V$ in the probability
density, see e.g \cite{Ak}. The related universality questions will
be addressed elsewhere \cite{prep}.  

YVF is grateful to Jac
Verbaarschot for stimulating his interest in the problem and useful
communications.  This work was supported by EPSRC grant GR/R13838/01
"Random Matrices close to unitary or Hermitian" (YVF) and by a
Heisenberg fellowship of the Deutsche Forschungsgemeinschaft (GA).


\begin{thebibliography}{99}
\bibitem{note} Note that the factor
$\left[\frac{\det{X_b}}{\det{X_f}}\right]^{\nu}$ is absent in the
corresponding expression in \cite{1}. Instead, it is (implicitly)
included in the normalisation of the probability measure of the
chiral random matrix ensemble.
\end{thebibliography}
\end{document}